\documentclass[amsmath,amssymb,nofootinbib,showpacs,twocolumn,nofootinbib]{revtex4-1}
\usepackage{dcolumn}
\usepackage{bm}
\usepackage{graphicx}
\usepackage{epstopdf}
\usepackage{subfig}
\usepackage{epsfig}
\usepackage{color}
\usepackage{wasysym}
\usepackage{natbib}
\usepackage{twoopt}
\usepackage{dashrule}
\usepackage{mathrsfs}
\usepackage{ulem}
\usepackage[breaklinks=true,colorlinks=true,linkcolor=blue,citecolor=blue,urlcolor=blue,pdfauthor={JieFeng},pdftitle={combinedFit}]{hyperref}

\definecolor{ColorTitle}{cmyk}{0,.88,.77,.40}

\newcommand{\AMS}{\textsf{AMS-02}}

\newcommand{\eg}{\textit{e.g.}}
\newcommand{\ie}{\textit{i.e.}}

\newcommand{\p}{\textsf{p}}
\newcommand{\He}{\textsf{He}}

\newcommand{\BC}{\textsf{B}/\textsf{C}}

\newcommand{\eplus}{\ensuremath{e^{+}}}
\newcommand{\eminus}{\ensuremath{e^{-}}}
\newcommand{\muplus}{\ensuremath{\mu^{+}}}
\newcommand{\muminus}{\ensuremath{\mu^{-}}}
\newcommand{\tauplus}{\ensuremath{\tau^{+}}}
\newcommand{\tauminus}{\ensuremath{\tau^{-}}}
\newcommand{\q}{\ensuremath{q}}
\newcommand{\qbar}{\ensuremath{\bar{q}}}
\newcommand{\bb}{\ensuremath{b}}
\newcommand{\bbar}{\ensuremath{\bar{b}}}
\newcommand{\W}{\ensuremath{W}}
\newcommand{\Wbar}{\ensuremath{\bar{W}}}

\newcommand{\epm}{\ensuremath{e^{\pm}}}
\newcommand{\pbarp}{\textsf{\ensuremath{\bar{p}/p}}}
\newcommand{\pbar}{\textsf{\ensuremath{\bar{p}}}}

\newcommand{\BeBe}{\ensuremath{{}^{10}\textsf{Be}{/}{}^{9}\textsf{Be}}}

\newcommand{\Dragon}{\texttt{DRAGON}}

\begin{document}
\title{Dark Matter Search in Space: Combined Analysis of Cosmic Ray Antiproton-to-Proton Flux Ratio and Positron Flux Measured by AMS-02}
\author{Jie Feng$^{1,2}$}
\author{Hong-Hao Zhang$^{1,}$}
\email{zhh98@mail.sysu.edu.cn}
\affiliation{$^1$School of Physics, Sun Yat-Sen University, Guangzhou 510275, China\\
$^2$Massachusetts Institute of Technology (MIT), Cambridge, Massachusetts 02139, USA}
\begin{abstract}
Dark matter search in space has been carried out for many years. Measurements of cosmic ray photons,
     charged antiparticles and neutrinos are useful tools for dark matter indirect search.
The antiparticle energy spectra of cosmic rays have several exciting features such as the
unexpected positron excess at $E\sim$ 10 -- 500\,GeV and the remarkably flattening antiproton/proton  at $E\sim$ 60--450\,GeV
precisely measured by the AMS-02 experiment, which can not be explained simultaneously by secondary production in
interstellar medium.
In this work, we report a combined analysis of cosmic ray antiproton and positron spectra arising from dark matter
on the top of a secondary production in a spatial-dependent propagation model.
We discuss the systematic uncertainties from antiproton
production cross section using the two latest Monte Carlo generators, \textit{i.e.} EPOS LHC and QGSJET-II-04m, respectively. We compare their results.
In the case of EPOS LHC, we find that the dark matter pair annihilating into $\tau$ leptons channel with 100\% branching ratio and p-wave annihilation cross section assumption is the only
possible one channel scenario to explain data.
On the other hand, there is not a single possible channel in the case of QGSJET-II-04m.
We also propose possible two-channel scenarios based on these two Monte Carlo generators.
\end{abstract}
\maketitle

%%%%%%%%%%%%%%%%%%%%%%%%%%%%%%
\section{Introduction}\label{Sec::Introduction} %%%
%%%%%%%%%%%%%%%%%%%%%%%%%%%%%%

After nearly one century of physics investigation, the search for dark matter is still ongoing.
This search is carried out in three complementary ways: dark matter production in colliders,
direct detection with underground instruments and indirect detection in cosmic rays (CRs).
Dark matter annihilation or decay may produce elementary particles, including neutral particles
(photons ($\gamma$) and neutrinos) and charged ones ( positrons (\eplus) and antiprotons (\pbar{}) ).
An impressive amount of dark matter information is being achieved by
$\gamma$-ray data coming from spacebased or groundbased telescopes such as Fermi's Large Area
Telescope (Fermi-LAT) \cite{FermiLAT:2012ApJ, Ackermann:2015zua} or High Energy Stereoscopic System (H.E.S.S)
	\cite{Abdallah:2016jja}.
Besides, valuable pieces of dark matter information from neutrinos are being collected
by IceCube \cite{Aartsen:2016pfc}.
At the same time, an increase in the accuracy of charged elementary CR particles spectra is driving us
to a deeper understanding of the fundamental physics processes in the Galaxy.
Thanks to the new generation detection experiments,
such as the Payload for Antimatter Matter Exploration and Light-nuclei Astrophysics (PAMELA) or
the Alpha Magnetic Spectrometer (\AMS{}) in space,
we are able to retrieve dark matter information in charged particle channels.
The \AMS{} collaboration has now published the precise \pbarp{} ratio measurement
between $\sim$\,0.5 and $\sim$\,450\,GeV of kinetic energy, showing that the
ratio above $\sim$\,60\,GeV experiences a remarkably flat behavior \cite{Aguilar:2016kjl}.
PAMELA has also published similar results but with less statistical significance \cite{Adriani:2012paa}.
Together with the resent \eplus{} flux data \cite{Adriani:2008zr,Aguilar:2014mma},
which shows a surprising excess above $\sim$\,10\,GeV, those results give us a hint of extra sources.

Unlike neutral particles that travel almost along straight lines,
charged particles are difficult to be traced back to their sources
due to the complex magnetic turbulence in the Galaxy.
To constrain secondary production contribution, one also need to study the CR \BC{} elemental ratio,
   which have been measured by PAMELA and \AMS{} in space, or by or
   the Advanced Thin Ionization Calorimeter (ATIC-2) and
   the Cosmic Ray Energetics and Mass (CREAM) on balloon.
Besides, systematics from solar modulation and antiparticle production cross section
should also be studied \cite{Feng:2016loc}. 
Recent studies \cite{Feng:2016loc,Boschini:2017fxq} 
showed that the excess of antiprotons was not significant but
that of positrons was solid given by the current understanding of systematics.
Some studies were carried out to interpret the positron excess that were consistent
with a smooth \BC{} spectrum. 
According to diffusive shock acceleration (DSA), the sources accelerating C-N-O are the same as those accelerating helium or protons, 
which are the main progenitors of antiprotons and positrons \cite{Blasi:2009hv, Mertsch:2014poa, Tomassetti:2015cva}.
However, a recent deuteron-to-helium ratio (d/He) measurement at 0.5-2 TeV/n by the satellite mission SOKOL \cite{TURUNDAEVSKIY2017496}  showed a rather high value, which is not expected from the predictions tuned against \BC{}. 
It stimulates a challenge to DSA \cite{Tomassetti:2016fsh}. 
If this deuteron-to-helium ratio measurement is correct, one should expect that 
the sources accelerating C-N-O are not the same as those accelerating helium or protons. 
The positron excess can be explained by nearby sources, which should be compatible with d/He instead of \BC{}.
Otherwise, 
it seems unavoidable to introduce extra source components
such as dark matter particle annihilation \cite{Cirelli:2008jk,Cirelli:2008pk, Yuan:2013eja,Lin:2014vja,Boudaud:2014dta},
or $e^{\pm}$ pair production mechanisms inside nearby pulsars
\cite{Hooper:2008kg,Profumo:2008ms,Delahaye:2010ji,Linden:2013mqa,Yin:2013vaa,Delahaye:2014osa,Lin:2014vja,Boudaud:2014dta, Feng:2015uta}.
Observations by Fermi-LAT \cite{Abdo:2011ApJ} indicated that $\gamma$-rays of pulsars
were produced by leptons rather than hadrons, which can basically exclude the possibility
that pulsars produce high energy antiprotons.

Numerous analyses have been performed to interprete precise \pbarp{} spectrum measured by AMS-02
independent of \eplus{} with dark matter scenarios
\cite{Giesen:2015ufa,Lin:2015taa,Lin:2016ezz,Huang:2016tfo,Cui:2016ppb,Cuoco:2016eej}.
There are also some combined analyses of PAMELA \pbar{}, which has larger uncertainties,
      and \AMS{} \eplus{} \cite{Cheng:2016slx}.
In this paper, we perform a combined analysis of \pbarp{} and \eplus{} in dark matter
scenarios.
We reduce some uncertainty from normalization by analyzing \pbarp{} instead of \pbar{} spectrum
because \p{} are \pbar{} progenitors.
For a similar reason, we avoid injection uncertainties of \eminus{} by analyzing
\eplus{} instead of \eplus{}/(\eplus{}+\eminus{}).
Our basic idea is that the cross section and the mass of dark matter annihilation estimated from \eplus{} data
should be consistent with that from \pbar{} data.
Besides, we notice that antiproton production cross section introduces major systematic uncertainties
in \pbarp{} spectrum \cite{Feng:2016loc, DiMauro:2014iia, Winkler:2017xor}. Following the implementation of the cross section
from MC generators in \cite{Feng:2016loc}, we present our study with EPOS LHC and QGSJET-II-04m, which were tuned
against the latest LHC experimental data and reproduce the \pbar{} production well \cite{Feng:2016loc,Lin:2016ezz}.
In each case, we perform a global fit to the data with all the free parameters in the propagation and dark matter models.
We quantify the agreement between model prediction and data with ``p-value'' method.
We find that $\chi\chi\rightarrow\tauplus\tauminus$ is the only possible channel,
 with 100\% branching ratio and p-wave cross section assumption,
based on the antiproton background calculated by EPOS LHC, while no channel is possible for QGSJET-II-04m.
We also study the scenarios that dark matter decays into two channels, which gives a larger p-value
compared to the one from channel scenarios.
Comparisons with the analyses of $\gamma$-ray and cosmic microwave background (CMB) observations are also shown.

This paper is organized as follows. In Sec.\ref{Sec::calculations},
we present our calculations. In Sec.\ref{Sec::Astro-physical_background},
we review briefly the \pbar{} and \eplus{} from astro-physical sources as
the background of our analysis. In Sec.\ref{Sec::anti-matter}, we introduce
\pbar{} and \eplus{} flux produced at dark matter annihilation.
In Sec.\ref{Sec::statistical_test}, it is presented our definition of a good fit.
In Sec.\ref{Sec::Solar_modulation}, our consideration of solar modulation uncertainties is shown.
In Sec.~\ref{Sec::results}, we show our results and discussion including one annihilation
channel in Sec.~\ref{Sec::one-channel} and two annihilation channels in Sec.~\ref{Sec::two-channels}.
Finally, the conclusion is drawn in Sec.~\ref{Sec::conclusion}.
%%%%%%%%%%%%%%%%%%%%%%%
\section{calculations}
\label{Sec::calculations}
%%%%%%%%%%%%%%%%%%%%%%%

%%%%%%%%%%%%%%%%%%%%%%%
\subsection{Astro-physical background}
\label{Sec::Astro-physical_background}
%%%%%%%%%%%%%%%%%%%%%%%
%%%%%%%%%%%%%%%%%%%%%%%%%%%%%%%%%%%%%%%%%%%%%%%%%%%%%%
\begin{figure}[!t]
\includegraphics[width=0.46\textwidth]{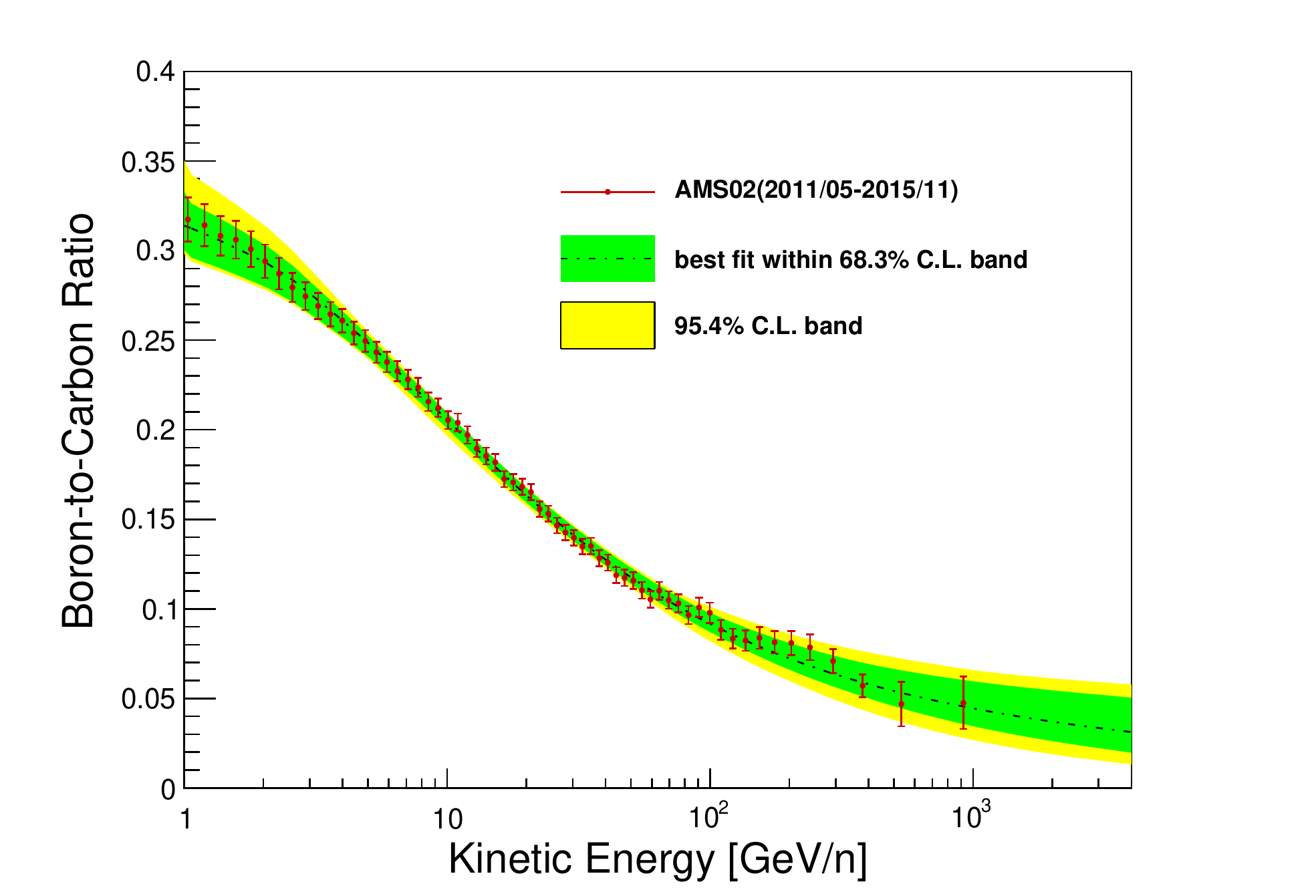}
\caption{
   Best fit model calculation and uncertainty band for the
B/C ratio in comparison with \AMS{} data \cite{Aguilar:2016vqr}.
}
\label{Fig::bc}
\end{figure}
%%%%%%%%%%%%%%%%%%%%%%%%%%%%%%%%%%%%%%%%%%%%%%%%%%%%%%%%
%%%%%%%%%%%%%%%%%%%%%%%%%%%%%%%%%%%%%%%%%%%%%%%%%%%%%%
\begin{figure}[!t]
\includegraphics[width=0.46\textwidth]{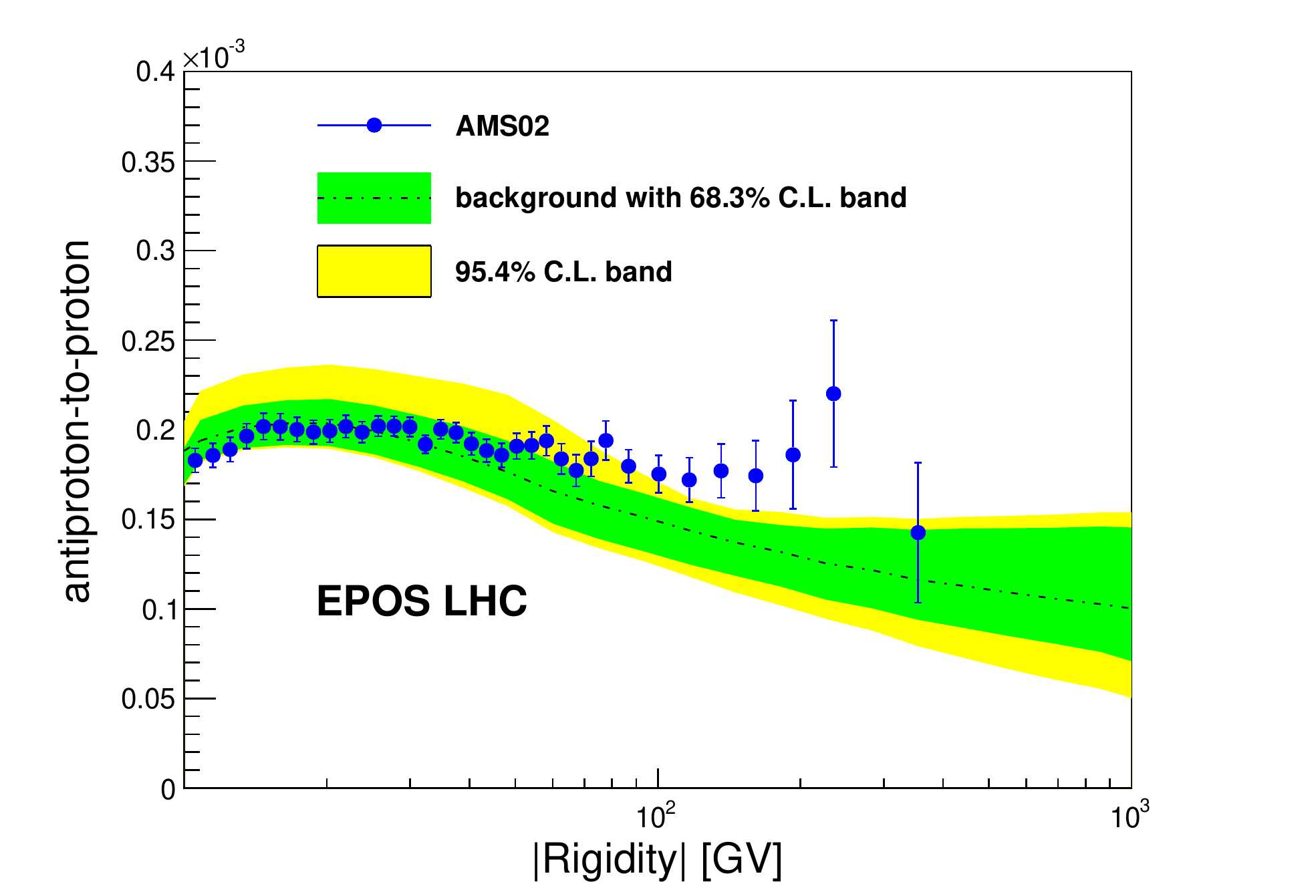}
\includegraphics[width=0.46\textwidth]{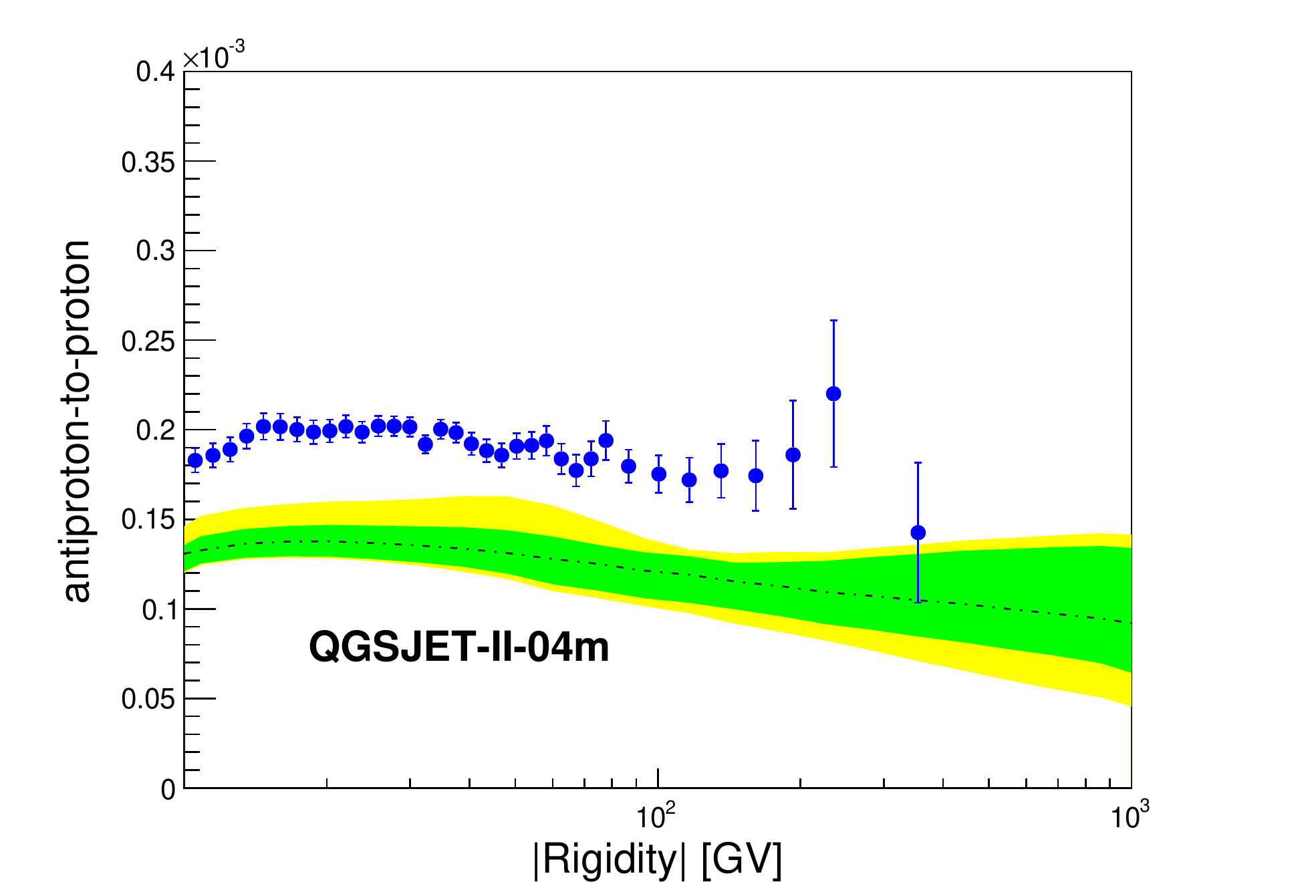}
\caption{
   Model prediction using the best fit parameters and uncertainty band for the antiproton/proton ratio.
   TOP: EPOS LHC hadronic model prediction.
   BOTTOM: QGSJET-II-04m hadronic model prediction.
   \AMS{} data \cite{Aguilar:2016kjl} is also shown for comparison.
}
\label{Fig::ap}
\end{figure}
%%%%%%%%%%%%%%%%%%%%%%%%%%%%%%%%%%%%%%%%%%%%%%%%%%%%%%

In convectional CR propagation models, antiparticles are only produced in collisions
of high-energy nuclei with interstellar medium (ISM). The fluxs of their progenitor nuclei and CR propagation
process together determine the specta of antiparticles.
The Galatic disk is surrounded by a halo with half-thickness $L$. For each CR species,
its propagation can be described by a two-dimensional transport equation:
\begin{equation}\label{Eq::DiffusionTransport}
	\frac{\partial {\psi}}{\partial t} = Q + \vec{\nabla}\cdot (D\vec{\nabla}{\psi}) - {{\psi}}{\Gamma} + \frac{\partial}{\partial E} (\dot{E} {\psi})  \,,
\end{equation}
where $\psi=\psi(E,r,z)$ is the number density as a function of energy and space coordinates,
$\Gamma= \beta c n \sigma$ is the destruction rate in ISM, with density $n$,
at velocity $\beta c$ and cross section $\sigma$. The source term $Q$ includes a primary term, $Q_{\rm pri}$,
and a secondary production term $Q_{\rm sec}= \sum_{\rm j} \Gamma_{j}^{\rm sp} \psi_{\rm j}$, from interaction of
heavier $j$--type nuclei with rate $\Gamma_{j}^{\rm sp}$.
The term $\dot{E}=\--\frac{dE}{dt}$ describes ionization and Coulomb losses, as well as radiative cooling of CR leptons.
The diffusion coefficient is taken as $D(p,z)=\beta D_0 (R/R_0)^{\delta(z)}$, where $D_0$ shows its normalization,
$R\equiv pc/Ze$ is defined as the magnetic rigidity and $R_0$ is its normalization rigidity.
$\delta(z)$ expresses the scaling index.

Recent studies were done to get
a set of injection and propagation parameters which could simultaneously reproduce
a large set of nuclear data including proton, helium and carbon fluxes, the \BC{} elemental
ratio, and the \BeBe{} isotopic ratio \cite{Lin:2014vja,Johannesson:2016,Feng:2016loc,Evoli:2015vaa}.
To assess astro-physical background of antiparticles, we adopt a spatial-dependent model of
CR diffusion \cite{Tomassetti:2012ga,Tomassetti:2015mha}.
This model explains the
high-energy departures from the standard universal power-law expectations
in \p{} and \He{} spectra observed by PAMELA \cite{Adriani:2011cu}
and confirmed by \AMS{} \cite{Aguilar:2015ooa, Aguilar:2015ctt},
predicts a harder secondary-to-primary flux ratio later observed by \AMS{}
and solves the problems on nuclei anisotropy and diffuse $\gamma$ rays \cite{Feng:2016loc, Guo:2018wyf}
while the convectional models failed to do so \cite{Evoli:2013lma}. In this scenario, the scaling
index $\delta(z)=\delta_0$ in the region of $|z|<\xi L$ (inner halo) and $\delta(z)=\delta_0+\Delta$
when $|z|\geq\xi L$ (outer halo) .
The normalization is $D_0$ for the inner halo and $\chi D_0$ for the outer.
There is a connecting function of the type $F(z)=(z/L)^n$ to ensure a smooth transition of the parameters $\chi$ and $\Delta$ across the two zones \cite{Guo:2015csa}.
The injection spectral indices of all the nuclei whose $z>1$ all equal to $\nu$, while that of proton is $\nu+\Delta\nu$.
Based on the method presented in Ref. \cite{Feng:2016loc}, we redo the Bayesian analysis on those parameters with the latest \AMS\ \BC\ ratio \cite{Aguilar:2016vqr}. 
 In Fig. \ref{Fig::bc}, the B/C ratio calculations are shown in comparisons
with the data.
We use \Dragon{} package \cite{Gaggero:2013rya}, which is based on GALPROP package \cite{Strong:2007nh}, to solve the transport equation.
In Table. \ref{Tab::bestfits}, we compare the fit parameters in the spatial dependent propagation model (this work) with those in the standard GALPROP model (SG) reported in \cite{Boschini:2017fxq}. 
At low rigidity, the diffusion coefficient has a velocity dependent factor $\beta^{\eta}$, where $\beta=v/c$. 
We fix $\eta=-4$ in order to reproduce proton and helium fluxes below 20 GV in this work, while it is a free parameter in Ref. \cite{Boschini:2017fxq}. This setting can avoid the complicated parameters associated to convection, reacceleration and the injection break around 7 GV. 
In the standard GALPROP model, the injection spectral index of protons or helium is no longer a constant and contains two breaks (\ie{} $R_1$ and $R_2$) with different indices (\ie{} $\nu_1$, $\nu_2$ and $\nu_3$) before and after the breaks.
We define $\nu_{1He} = \nu$ and $\nu_{1p} = \nu + \Delta\nu$ in order to compare them with those in Ref. \cite{Boschini:2017fxq}.
$V_{alf}$, $V_{conv}$ and $dV_{conv}/dz$, in Table. \ref{Tab::bestfits}, are the Alfv$\acute{e}$n velocity,  the convection wind velocity and its gradient, respectively.

%%%%%%%%%%%%%%%%%%%%%%%%
\begin{table}[!t]
  \begin{tabular}{c c c c c c c c c}
    \hline\hline
    \textbf{parameter} & $\,$\textbf{unit}$\,$ & $\,$\textbf{this work}$\,$ & $\,$\textbf{SG} $\,$\tabularnewline
    \hline\hline
    $\eta$ & \dots  &  -4  & 0.91\tabularnewline
    \hline
    $L$ & kpc  &  6.70  & 4.0\tabularnewline
    \hline
    $D_{0}$ & 10$^{28}$\,cm$^{2}$\,s$^{-1}$  & 2.18 & 4.3\tabularnewline
    \hline
    $\delta$ & \dots & 0.19 &  0.395 \tabularnewline
    \hline
    $\Delta$ & \dots & 0.56 & \dots \tabularnewline
     \hline
    $\xi$ & \dots & 0.22 & \dots \tabularnewline
    \hline
    $\chi$ & \dots & 0.30 & \dots \tabularnewline
      \hline
    $V_{alf}$ & km\,s$^{-1}$ & \dots & 28.6 \tabularnewline
      \hline
    $V_{conv}$ & km\,s$^{-1}$ & \dots & 12.4 \tabularnewline
      \hline
    $dV_{conv}/dz$ & km\,s$^{-1}$\,kpc$^{-1}$ & \dots & 10.2 \tabularnewline
       \hline
    $R_{1p}$ & GV & \dots &  7 \tabularnewline
       \hline
    $R_{2p}$ & GV & \dots &  360 \tabularnewline
       \hline
    $\nu_{1p}$ & \dots & 2.39 &  1.69 \tabularnewline
       \hline
    $\nu_{2p}$ & \dots & \dots &  2.44 \tabularnewline
       \hline
    $\nu_{3p}$ & \dots & \dots &  2.28 \tabularnewline
       \hline
    $R_{1p}$ & GV & \dots &  7 \tabularnewline
           \hline
    $R_{2p}$ & GV & \dots &  360 \tabularnewline
      \hline
      $\nu_{1He}$ & \dots & 2.29 & 1.71 \tabularnewline
         \hline
      $\nu_{2He}$ & \dots & \dots & 2.38 \tabularnewline
         \hline
      $\nu_{3He}$ & \dots & \dots & 2.21 \tabularnewline
      \hline\hline
  \end{tabular}
  \caption{\captionsize%
    Results of the MCMC scan for the transport and injection parameters in terms of best-fit values in the spatial-dependent propagation model (this work), compared with those in the standard GALPROP model (SG) \cite{Boschini:2017fxq}.}
  \label{Tab::bestfits}%
\end{table}
%%%%%%%%%%%%%%%%%%%%%%%%%%%%%%%%%%%%%%%%%%%%%%%%%%%%%%%%%%%%%%%%%%%%%%%%%%%%

Antiproton production cross section systematic is one of the main uncertainties of the astro-physical background.
As has been studied in \cite{Feng:2016loc, Kachelriess:2015wpa}, two of the most advanced Monte Carlo (MC) generators
EPOS LHC \cite{Pierog:2013ria} and QGSJET-II-04m \cite{Kachelriess:2015wpa} can reproduce the recent ground
experiments well.
However, due to the scarcity of the anti-neutron production data, we have no way to test anti-neutron production
cross sections. EPOS LHC predicts the anti-neutron/anti-proton ratio varies between 1.2 and 2.0, while
QGSJET-II-04m shows it is close to 1 except near the production threshold.
As is shown in Fig. \ref{Fig::ap}, both model predictions with the latest \AMS{}  \BC{} data \cite{Aguilar:2016vqr} on the antiproton-to-proton ratio
are 
below 
the experimental data measured by the same instrument \cite{Aguilar:2016kjl}.
The first measurement of antiproton production cross section in p + He $\rightarrow$ \pbar{} + X channel is recently made by LHCb experiment located at the Large Hadron Collider accelerator (LHC) at CERN \cite{Graziani:2017xas}. 
Collisions of 6.5 TeV proton beams on He Nuclei at rest have been studied. 
Preliminary results showed that the data were between the predictions of EPOS LHC and QGSJET-II-04m. 
One should also note that those measurements are focused on high transverse momentum ($p_{T}$) range, which is the tail of the production.
More data at low $p_{T}$, where most of the antiprotons are produced, will be appreciated.
We believe the truth should be
somewhere between these two models, so we test the dark matter scenarios with the backgrounds predicted by them
individually.
Positron production cross section is taken from a recent parameterization \cite{Kamae:2006bf}.
As you can find latter in Sec.~\ref{Sec::results}, positron production cross section is not a dominating component
of the total uncertainties, since the excess of \eplus{} from the background is significant.
So we do not discuss other \eplus{} cross section in this paper.

%pulsar do not produce pbar.
Pulsars are also important sources which produce secondary positrons. Previous studies showed that it is better to
explain positron spectrum with pulsar models rather than with dark matter models \cite{Boudaud:2014dta}.
This is ascribed to the fact that the profile of pulsar model usually has more degree of freedom than that of dark matter model.
For example, it is unavoidable to introduce at least three parameters in the pulsar fit, \ie, the cutoff energy,
the injection spectral index and the normalization \cite{Feng:2015uta}.
In dark matter scenarios, however, there are only two free parameters: the mass of the dark matter $m_\chi$ and the
normalization (\ie{} thermally averaged annihilation cross section
$\langle\sigma v\rangle$ in the case of annihilation, where $\sigma$ is the annihilation cross section and $v$ is
the velocity, or $\tau$ which is the lifetime in the case of decay \cite{Giesen:2015ufa}).
The $\gamma$-ray spectrum of a single pulsar is preferred to be explained by a leptonic model rather than hadronic one
\cite{Abdo:2011ApJ}.
The spectral index of $\gamma$-ray produced from pion-decay emission \cite{Ellison:2007bg} of hadronic interactions
should be harder than that through the Inverse-Compton scattering by leptons in a pulsar.
Observation of RX J1713.7-3946 supports the latter one.
One might easily explain the CR antiproton spectrum by dark matter and the positron by pulsar. In this way, there will be
five free parameters so everything can be explained. However, this is not what we are going to do in this paper. Since
the parameters of pulsars are not easy to be constrained, we do not consider contribution of them into the astro-physical
background.
%

%%%%%%%%%%%%%%%%%%%%%%%%%%%%%%%%%%%%%
\subsection{The fluxes of anti-matter from dark matter annihilation}
\label{Sec::anti-matter}
%%%%%%%%%%%%%%%%%%%%%%%%%%%%%%%%%%%%%
%%%%%%%%%%%%%%%%%%%%%%%%%%%%%%%%%%%%%%%%%%%%%%%%%%%%%%%
\begin{figure}[!t]
\includegraphics[width=0.46\textwidth]{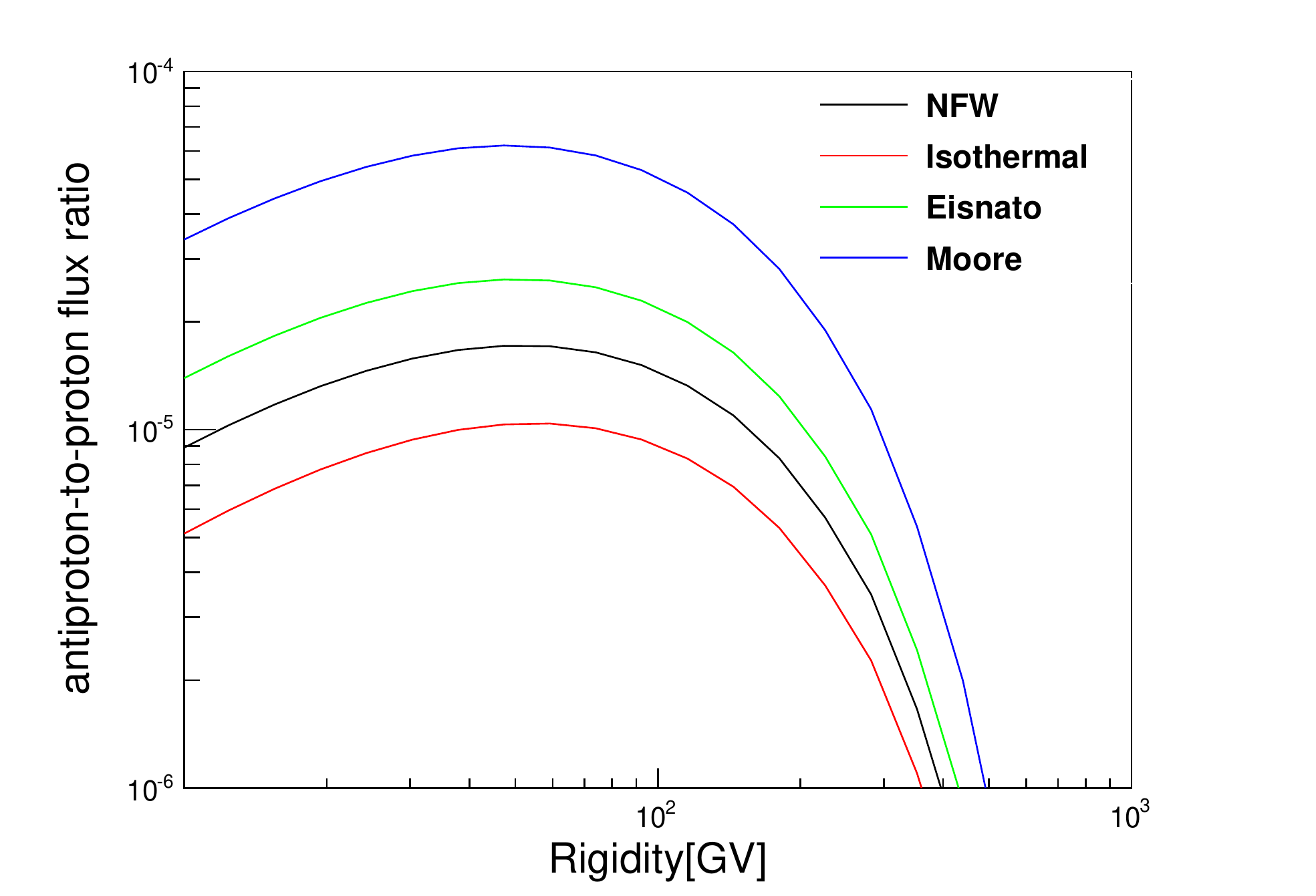}
\includegraphics[width=0.46\textwidth]{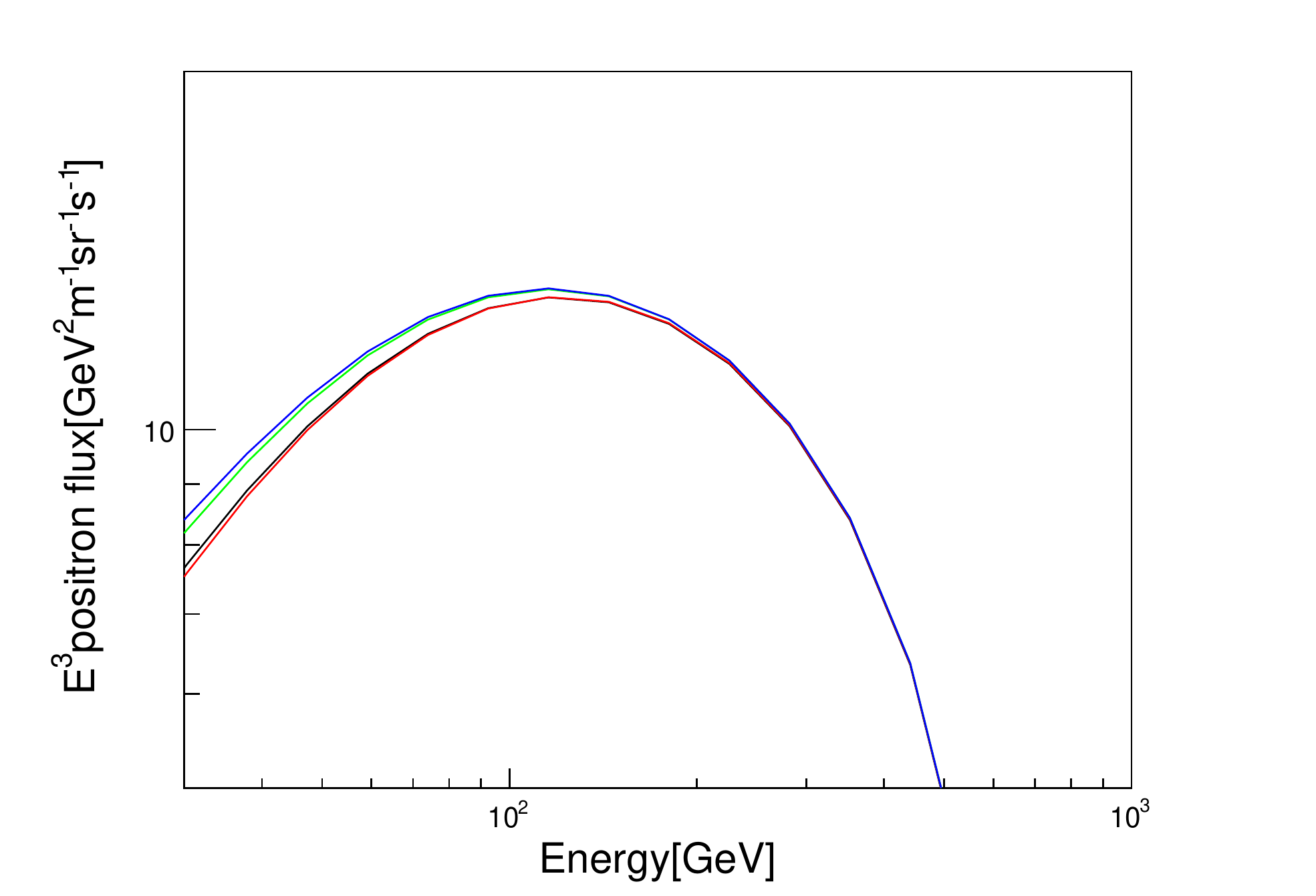}
\caption{
	Effects from dark matter distribution profiles on the observed spectra of \pbarp{} (TOP) and \eplus{} (BOTTOM). These plots are produced by the best fit parameters in Sec. \ref{Sec::one-channel}. 
}
\label{Fig::dm_profile}
\end{figure}
%%%%%%%%%%%%%%%%%%%%%%%%%%%%%%%%%%%%%%%%%%%%%%%%%%%%%%%
The CR anti-particle fluxes produced by dark matter have been studied and collected in A Poor Particle Physicist Cookbook
for Dark Matter Indirect Detection (PPPC) \cite{Cirelli:2010xx}. The authors calculated the results with PYTHIA and HERWIG
Monte Carlos so they had a feeling of the uncertainties.
Historically, leptons and vector bosons were treated as unpolarized. And parton showers were assumed not to emite $W$'s and
$Z$'s. Under these assumptions, \pbar{} will not be produced in leptonic channels. However, as is pointed out by
\cite{Ciafaloni:2010ti}, polarizations and electroweak corrections should be considered, which will modify \epm{} spectra at low
energies $E << m_\chi$ and produce \pbar{} in leptonic channels due to $W/Z$ radiation.

We consider dark matter annihilation into the following primary channels:
\eplus{}\eminus, muons (\muplus{}\muminus), tauons (\tauplus{}\tauminus), light quarks (\q\qbar),
bottom quarks (\bb\bbar{}) and W bosons (\W\Wbar{}) in order to compare this study with the $\gamma$-ray observations \cite{Ackermann:2015zua}.
As you can see latter in this paper, \q\qbar, \bb\bbar{} and \W\Wbar{} predict too many \pbar{} but not enough \eplus{}.
In order to improve the model, we also study the $VV\rightarrow4e$,  $VV\rightarrow4\mu$ and  $VV\rightarrow4\tau$,
where the annihilation first goes into a
new light boson $V$ that will later decay into a pair of leptons proposed by \cite{ArkaniHamed:2008qn, Pospelov:2008jd}.
Previous study by Ref.\cite{Cholis:2013lwa} showed that those channels can also reproduce \eplus{}.
These so-called ``4-body'' channels will not produce \pbar.
A recent study proposed a ``3-body'' channel where dark matter decays into a stable neutral particle and a pair of super symmetry fermions \cite{Cheng:2016slx},
  which is also interesting but more complex.
Another recent work proposed a new ``4-body'' channel that dark matter annihilate into light mediators that later decays into 2\q\qbar{} \cite{Huang:2016tfo}.
We do not discuss this case since it produces mostly \pbar{} in its final state while we prefer more \eplus{} in this study.

We adopt the Navarro, Frenk and White (NFW) \cite{Navarro:1995iw}
profile to describe the galactic distribution of dark matter in the Milky Way, which reads:
\begin{equation}\label{Eq::NFWProfile}
        \rho_\mathrm{NFW}\left(r\right) = \rho_{s}\frac{r_{s}}{r}\left(1+\frac{r}{r_{s}}\right)^{-2}  \,,
\end{equation}
where $\rho_{s}=0.184$ $GeV/cm^3$ and $r_{s}=24.42$ kpc are typical scale density and radius \cite{Cirelli:2010xx}. These
values are obtained by setting the density to be $\rho_{\odot}=0.3$ $GeV/cm^3$ at the Sun position $r_{\odot}=8.33$ kpc.
As is shown in Fig. \ref{Fig::dm_profile} , 
the dark matter density profile does not affect the observed positron spectrum near earth,
which is dominated by the local contribution.
It is also worth pointing out that the isothermal \cite{Bergstrom:1997fj}, Einasto \cite{Einasto:2009zd} and Moore \cite{Moore:1999nt, Diemand:2004wh} profile will change the dark matter antiproton contribution normalization by
a factor of $\sim$0.5, that of $\sim$2 and that of $\sim$4 respectively in the spatial-dependent propagation model.
These difference is smaller than those reported in traditional propagation models \cite{Jin:2014ica}.

After getting the fluxes of antiparticles produced by dark matter with NFW distribution, we take it as the source term in
the transport equation eq.~\eqref{Eq::DiffusionTransport}.
The differential fluxes of antimatter at production are $Q_\textsf{DM}(E)\propto (\rho/m_\chi)^{2}$ in
the case of annihilation and $Q_\textsf{DM}(E)\propto (\rho/m_\chi)$ in the case of decay \cite{Cirelli:2010xx}.
Since the fluxes have the same energy dependence for the two cases and
the energy spectra at the position of the earth would be similar,
we discuss dark matter annihilation here as an example.

%%%%%%%%%%%%%%%%%%%%%%%%%%%%%%%%%%%%%
\subsection{Formalism of the statistical test method}
\label{Sec::statistical_test}
%%%%%%%%%%%%%%%%%%%%%%%%%%%%%%%%%%%%%
We adopt a frequentist statistical test in this work.
Generally speaking, for discovering dark matter, we define the null hypothesis, $H_0$, as the astrophysical background, which is to be tested
against the alternative $H_1$ that includes both astrophysical background and dark matter signal.
For setting dark matter limits, we define $H_0$ as the astrophysical background plus dark matter signal to be tested against the
background-only hypothesis, $H_1$.
This work is in the former case.
To quantify the agreement between data and the predictions of $H$, we compute the
probability, the widely used ``p-value'' \cite{Cowan:2011epj},
\begin{equation}\label{Eq::p-value}
        p_\theta = \int^{\infty}_{t_{\theta,obs}}f\left(t_{\theta}|\theta\right) dt_{\theta}  \,,
\end{equation}
where $t_{\theta}$ is the $\chi^2$ for a given signal
strength $\theta$. $t_{\theta,obs}$ is the observed one.
$f\left(t_{\theta}|\theta\right)=\frac{t_\theta^{r/2-1}}{\Gamma(\frac{1}{2}r)2^{r/2}}e^{-t_\theta/2}$
is the distribution of $t_\theta$ for the number of degree of freedom $r$, where $\Gamma(x)$ is a gamma function.
%%%%%%%%%%%%%%%%%%%%%%%%%%%%%%%%%%%%%
\subsection{Solar modulation uncertainties}
\label{Sec::Solar_modulation}
%%%%%%%%%%%%%%%%%%%%%%%%%%%%%%%%%%%%%
Force field approximation is used to describe solar modulation.
However, this approximation fails to describe charge-sign-dependent solar modulation \cite{Maccione:2012cu, Kappl:2015hxv},
which is recently observed by PAMELA \cite{Adriani:2016uhu} and can be quantitatively studied with high statistic AMS data.
In order to take solar modulation uncertainties into account, the {$\chi^2$} can be written as a function of
\textbf{$\langle\sigma v\rangle$} and the dark matter mass {$m_\chi$},
\begin{equation}\label{Eq::chisquare2}
\chi^2\left(m_{\chi},\langle\sigma v\rangle,\bm{\theta_{bkg}},\phi\right)=
\sum_{k=1}^{N_{D}} \left( \frac{ y_{k}^{\rm exp} -
y_{k}^{\rm th}({m_{\chi},\langle\sigma v\rangle,\bm{\theta_{bkg}},\phi})}{\sigma_{k}} \right)^2,
\end{equation}
where $\sigma_{k}=\sqrt{\sigma_{k,0}^2+\sigma_{k,\phi}^2}$ is the total uncertainty of the data point $k$
with the model uncertainty ($\sigma_{k,\phi}$) introduced by varying the solar modulation potential $\phi$
from $-$300MV to $+$700MV.
The prior of background parameters $\bm{\theta_{bkg}}$
is obtained via the fitting to the B/C \cite{Aguilar:2016vqr},
$^{10}$Be$/^9$Be \cite{Hagen:1977ApJ,Buffington:1978ApJ,Webber:1979ICRC,Garcia-Munoz:1977ApJ,Garcia-Munoz:1981ICRC,Wiedenbeck:1980ApJ},
proton \cite{Aguilar:2015ooa}, helium \cite{Aguilar:2015ctt,Yoon:2011aa}
and carbon data \cite{Adriani:2014xoa,Ahn:2009tb}.
This quantity describes the consistency of model parameters ($m_{\chi},\langle\sigma v\rangle,\bm{\theta_{bkg}},\phi$)
and experimental data ($y^{\rm exp}$) with corresponding uncertainties ($\sigma_{k,0}$).

In this way, the model is more sensitive to high energy CR data rather than low energy ones. This method allows us to make use of the low energy data without introducing bias from solar modulation models.

%%%%%%%%%%%%%%%%%%%%%%%%%%%%%%%%%%%%%

%%%%%%%%%%%%%%%%%%%%%%%%%%%%%%%%%%%%%
\section{results and discussion}
\label{Sec::results}
%%%%%%%%%%%%%%%%%%%%%%%%%%%%%%%%%%%%%

%%%%%%%%%%%%%%%%%%%%%%%%%%%%%%%%%%%%%
\subsection{dark matter annihilation into one channel}
\label{Sec::one-channel}
%%%%%%%%%%%%%%%%%%%%%%%%%%%%%%%%%%%%%
%%%%%%%%%%%%%%%%%%%%%%%%%%%%%%%%%%%%%%%%%%%%%%%%%%%%%%%
\begin{figure}[!t]
\includegraphics[width=0.46\textwidth]{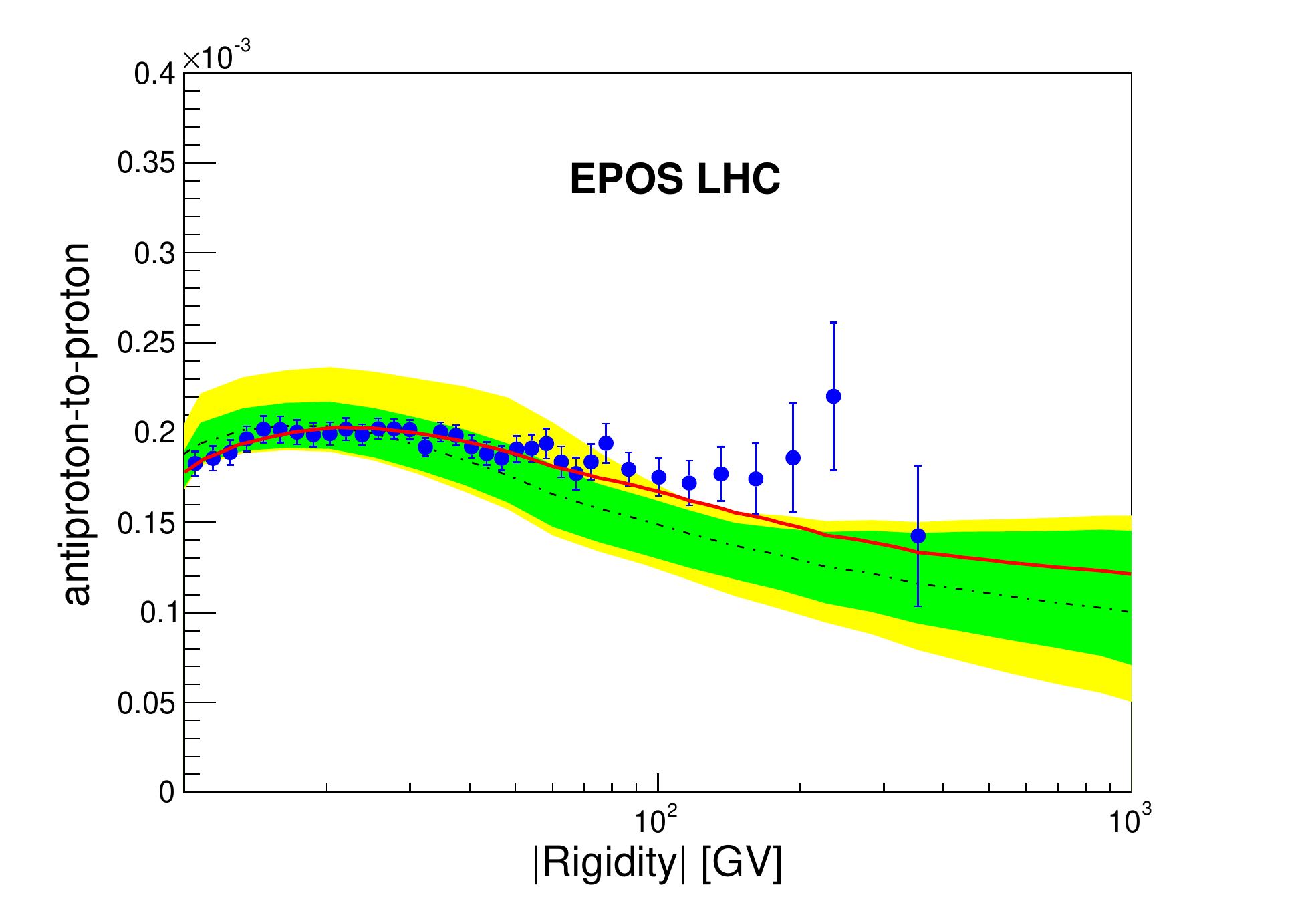}
\includegraphics[width=0.46\textwidth]{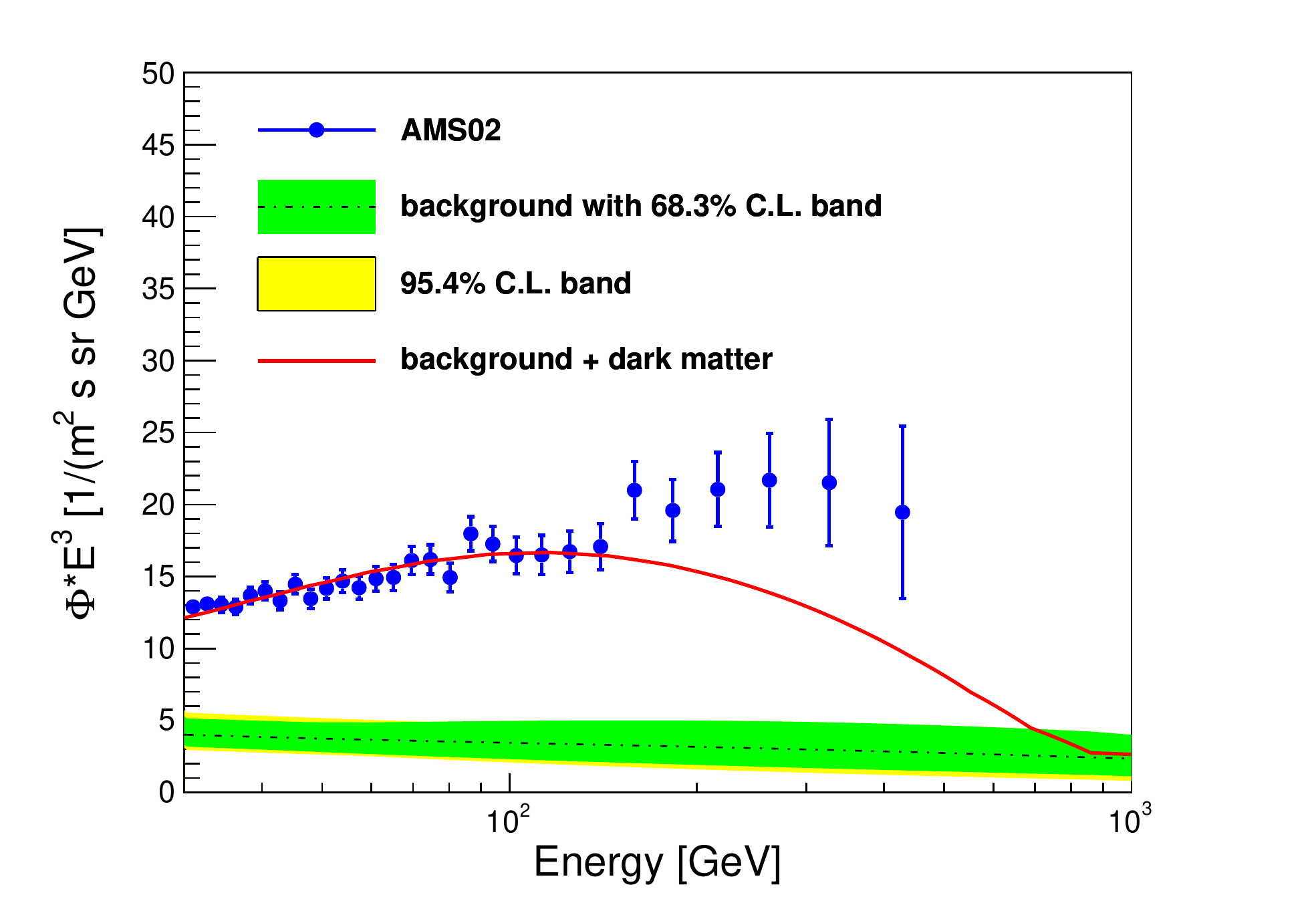}
\caption{
	Combined fit of \pbarp{} (TOP) and \eplus{} (BOTTOM) for the 1-channel scenario with EPOS LHC: $\chi\chi\rightarrow\tauplus\tauminus$.
}
\label{Fig::fit_1channel}
\end{figure}
%%%%%%%%%%%%%%%%%%%%%%%%%%%%%%%%%%%%%%%%%%%%%%%%%%%%%%%
%%%%%%%%%%%%%%%%%%%%%%%%%%%%%%%%%%%%%%%%%%%%%%%%%%%%%%%
\begin{figure}[!t]
\includegraphics[width=0.46\textwidth]{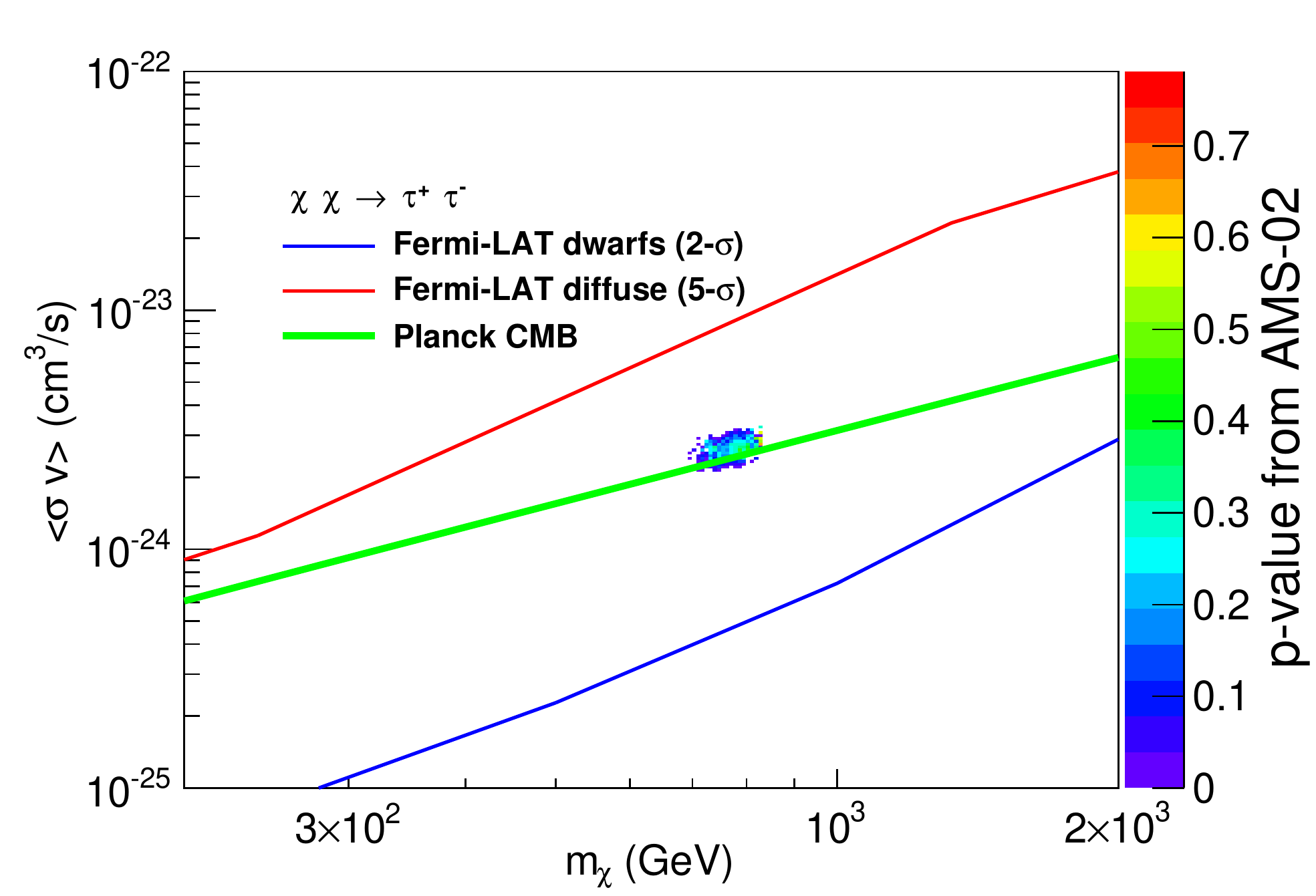}
\caption{
AMS-02 estimation of dark matter mass $m_\chi$ for \tauplus\tauminus{} channel with EPOS LHC,
comparing with
constraints from Fermi-LAT diffuse measurements \cite{FermiLAT:2012ApJ} in red, 6-year Dwarft Spheroidal Galaxies observation \cite{Ackermann:2015zua} in blue and Planck CMB observation \cite{Slatyer:2015jla, Ade:2015xua}.
}
\label{Fig::FermiLAT}
\end{figure}
%%%%%%%%%%%%%%%%%%%%%%%%%%%%%%%%%%%%%%%%%%%%%%%%%%%%%%%

We investigate the possibility to explain \pbar{} and \eplus{} by one annihilation channel with 100\% branching ratio.
We study \pbarp{} spectrum instead of \pbar{} flux, since the uncertainties of \pbar{} and those of $p$ are cancelled.
To avoid the uncertainties of the \eminus{} injection spectra, we study \eplus{} flux.
To avoid solar modulation uncertainties, we use positron flux data above 30 GeV, antiproton-to-proton flux ratio above 10 GeV,
and primary proton and nuclei fluxes above 10 GeV.
We found that only $\chi\chi\rightarrow\tauplus\tauminus$ channel gives us a p-value greater than $10^{-5}$, with a
normalized chisquare $\chi^{2}/n.d.f.=161.82/207$ and $p_{m_\chi}$=0.9918.
We get the best fit values: $m_\chi$ = $783\pm56$ GeV and
\textbf{$\langle\sigma v\rangle$}=$261.20\pm23.93\times 10^{-26}cm^3/s$.
Other channels are impossible.
To give a feeling of the goodness of the fit, we plot the calculated \pbarp{} and \eplus{}
spectra together with the AMS-02 data in Fig.~\ref{Fig::fit_1channel}.
On the other hand, however, there is no channel that gives us a large p-value with QGSJET-II-04m.

In Fig.~\ref{Fig::FermiLAT}, it is shown that this scenario survives from the constraints of Fermi-LAT diffuse measurements \cite{FermiLAT:2012ApJ}, 
has an overlap with Planck CMB constrain \cite{Slatyer:2015jla, Ade:2015xua}, but has been excluded by $\gamma$-ray
observations from Milky Way Dwarf Spheroidal Galaxies under the s-wave-dominated dark matter assumption \cite{Ackermann:2015zua}.
A latest study by \cite{Masi:2015fca} also pointed out that dark matter scenario obtained from CRs positrons are not completely excluded by CMB observations considering the current systematic uncertainties. 
One should notice that there is still a possibility to accept this scenario
if p-wave annihilation is not negligible, according to a recent study \cite{Zhao:2016xie}.

From this exercise, it is shown that most of the single channel scenarios can not simultaneously explain \pbar{} and \eplus{}.
With respect to the astro-physical background, the excess of \eplus{} is a solid evidence of
extra \eplus{} source, while that of \pbar{} is marginal. This requires a large
\textbf{$\langle\sigma v\rangle$} to explain \eplus{} data,
while a small \textbf{$\langle\sigma v\rangle$} to produce \pbar{}. For quark (\eg{} \q\qbar{} and \bb\bbar{})
or boson channels (\eg{} \W\Wbar{}), it predicts not enough \eplus{} and too much \pbar{}.
For leptonic channels (\ie{} \eplus{}\eminus{}, \muplus{}\muminus{} and \tauplus{}\tauminus{}),
    it predicts enough \eplus{} but the
\eplus{} profile of dark matter signal does not match data quite well.
Thus, we introduce one more channel to get more \eplus{} in Sec.~\ref{Sec::two-channels} to improve the fit.
%%%%%%%%%%%%%%%%%%%%%%%%%%%%%%%%%%%%%
\subsection{dark matter annihilation into two channels\label{Sec::two-channels}}%
%%%%%%%%%%%%%%%%%%%%%%%%%%%%%%%%%%%%%
%%%%%%%%%%%%%%%%%%%%%%%%%%%%%%%%%%%%%%%%%%%%%%%%%%%%%%%%%%%%%%%%%%%%%%%%%
\begin{table*}[!t]
\renewcommand{\arraystretch}{1.2}
\begin{tabular}{ c c c c c c c c}
\hline\hline

\textbf{$CH_1$} &\textbf{$CH_2$} & \textbf{$m_{\chi}$} & \textbf{$\langle\sigma v\rangle\times BR_1$} & \textbf{$\langle\sigma v\rangle\times BR_2$} &
\textbf{$\chi^2$} & \textbf{p-value}
\tabularnewline
 & & (GeV) & ($10^{-26}cm^3/s$) & ($10^{-26}cm^3/s$) & &
\tabularnewline
\hline
\hline
\tauplus\tauminus{} & VV$\rightarrow$2\tauplus\tauminus{} &
$1320\pm14$ &  $225.29\pm7.39$ &  $244.48\pm9.38$ & 164.42 & 0.9885 \tabularnewline
\hline
\q\qbar{}           & VV$\rightarrow$2\muplus\muminus{}  &
$654\pm12$ &  $0.96\pm0.14$ &  $89.85\pm5.7$ & 174.85 & 0.9593 \tabularnewline
\hline
\q\qbar{}           & VV$\rightarrow$2\tauplus\tauminus{} &
$1800\pm23$ &  $3.97\pm0.08$ &  $588.71\pm12.53$ & 162.43 & 0.9916 \tabularnewline
\hline
\bb\bbar{}          & VV$\rightarrow$2\muplus\muminus{}   &
$601\pm12$ &  $1.00\pm0.16$ &  $81.89\pm5.46$ & 185.60 & 0.8659 \tabularnewline
\hline
\bb\bbar{}          & VV$\rightarrow$2\tauplus\tauminus{} &
$1679\pm34$ &  $4.38\pm0.16$ &  $598.98\pm15.94$ & 149.29 & 0.9992 \tabularnewline
\hline
\W\Wbar{}           & VV$\rightarrow$2\muplus\muminus{}   &
$624\pm11$ &  $1.45\pm0.16$ &  $87.59\pm6.59$ & 180.12 & 0.9192 \tabularnewline
\hline
\W\Wbar{}           & VV$\rightarrow$2\tauplus\tauminus{} &
$1689\pm23$ &  $5.34\pm0.10$ &  $594.68\pm17.96$ & 143.65 & 0.9998 \tabularnewline

\hline\hline
\end{tabular}
\caption{\captionsize%
	$m_\chi s$, \textbf{$\langle\sigma v\rangle$s}, $\chi^2 s$ and p-values given by the ``Good'' fits in two-channel scenarios with EPOS LHC. The number of degree of freedom is 208.}
\label{Tab:results_eposlhc}%
\end{table*}
%%%%%%%%%%%%%%%%%%%%%%%%%%%%%%%%%%%%%%%%%%%%%%%%%%%%%%%%%%%%%%%%%%%%%%%%%%%%
%%%%%%%%%%%%%%%%%%%%%%%%%%%%%%%%%%%%%%%%%%%%%%%%%%%%%%%
\begin{figure}[!t]
\includegraphics[width=0.46\textwidth]{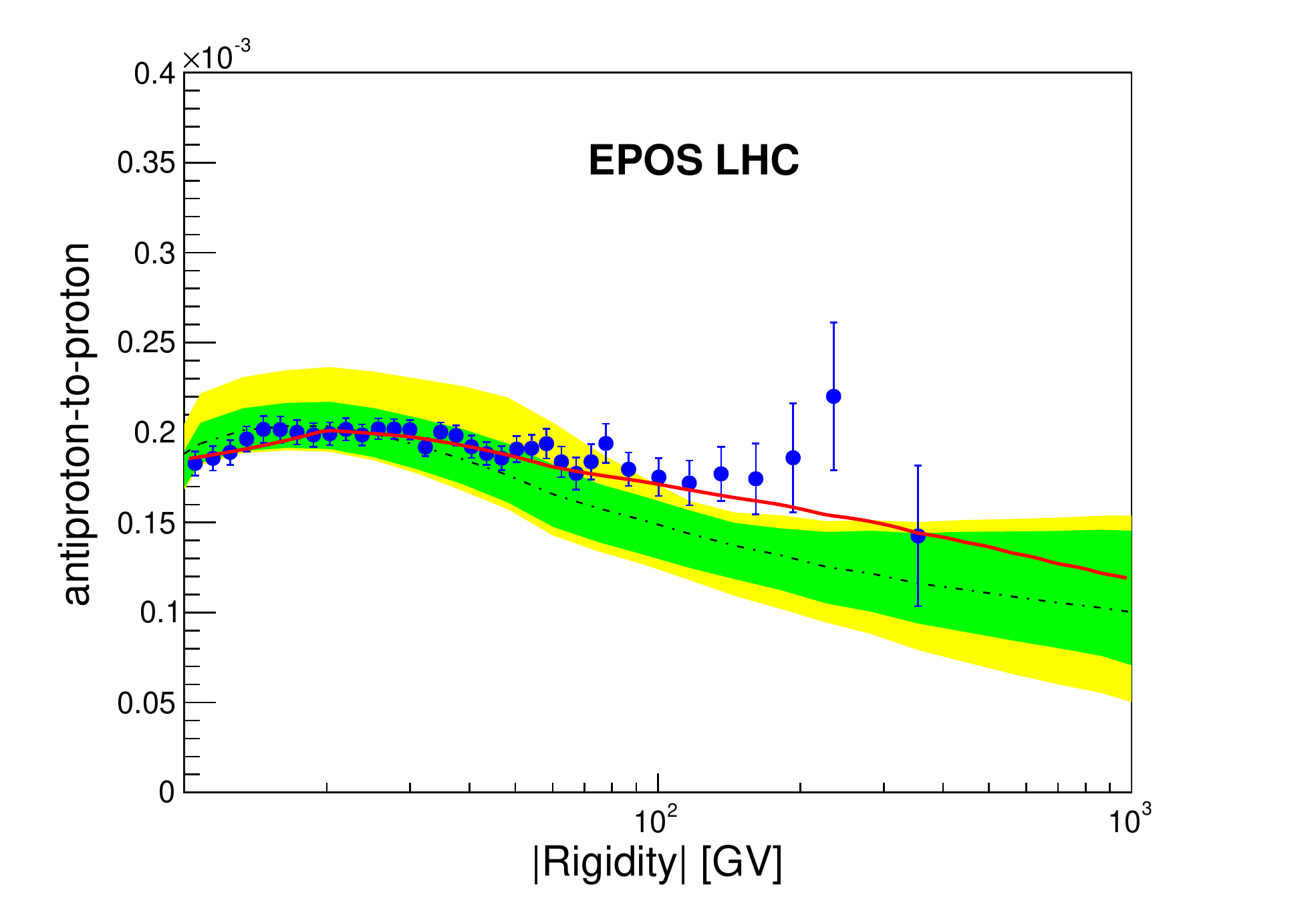}
\includegraphics[width=0.46\textwidth]{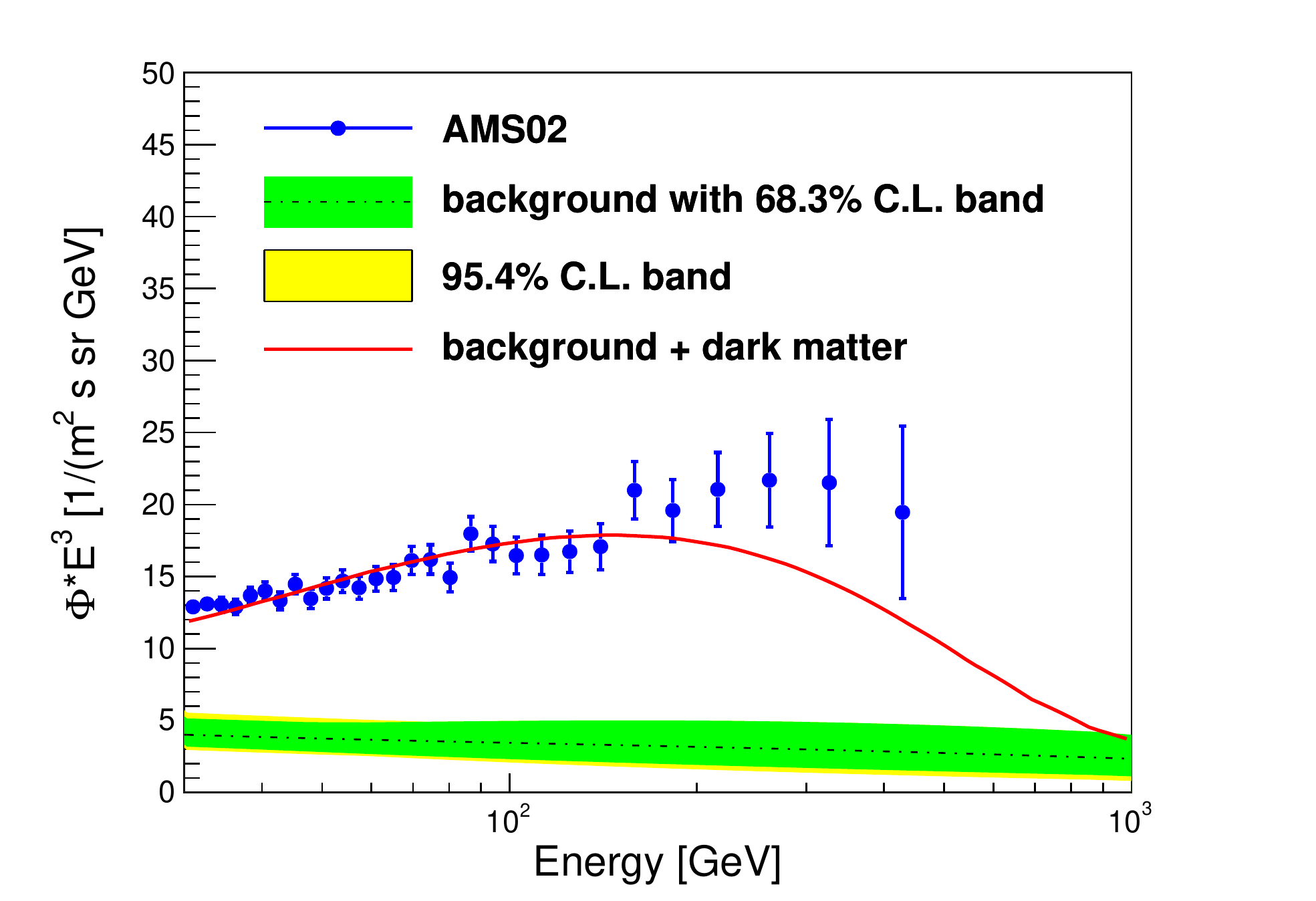}
\caption{
	Combined fit of \pbarp{} (TOP) and \eplus{} (BOTTOM) for the 2-channel scenario with EPOS LHC: $\chi\chi\rightarrow\W\Wbar$ and $\chi\chi\rightarrow VV \rightarrow2\tauplus\tauminus$.
}
\label{Fig::fit_2channel}
\end{figure}
%%%%%%%%%%%%%%%%%%%%%%%%%%%%%%%%%%%%%%%%%%%%%%%%%%%%%%%
%%%%%%%%%%%%%%%%%%%%%%%%%%%%%%%%%%%%%%%%%%%%%%%%%%%%%%%
\begin{figure}[!t]
\includegraphics[width=0.46\textwidth]{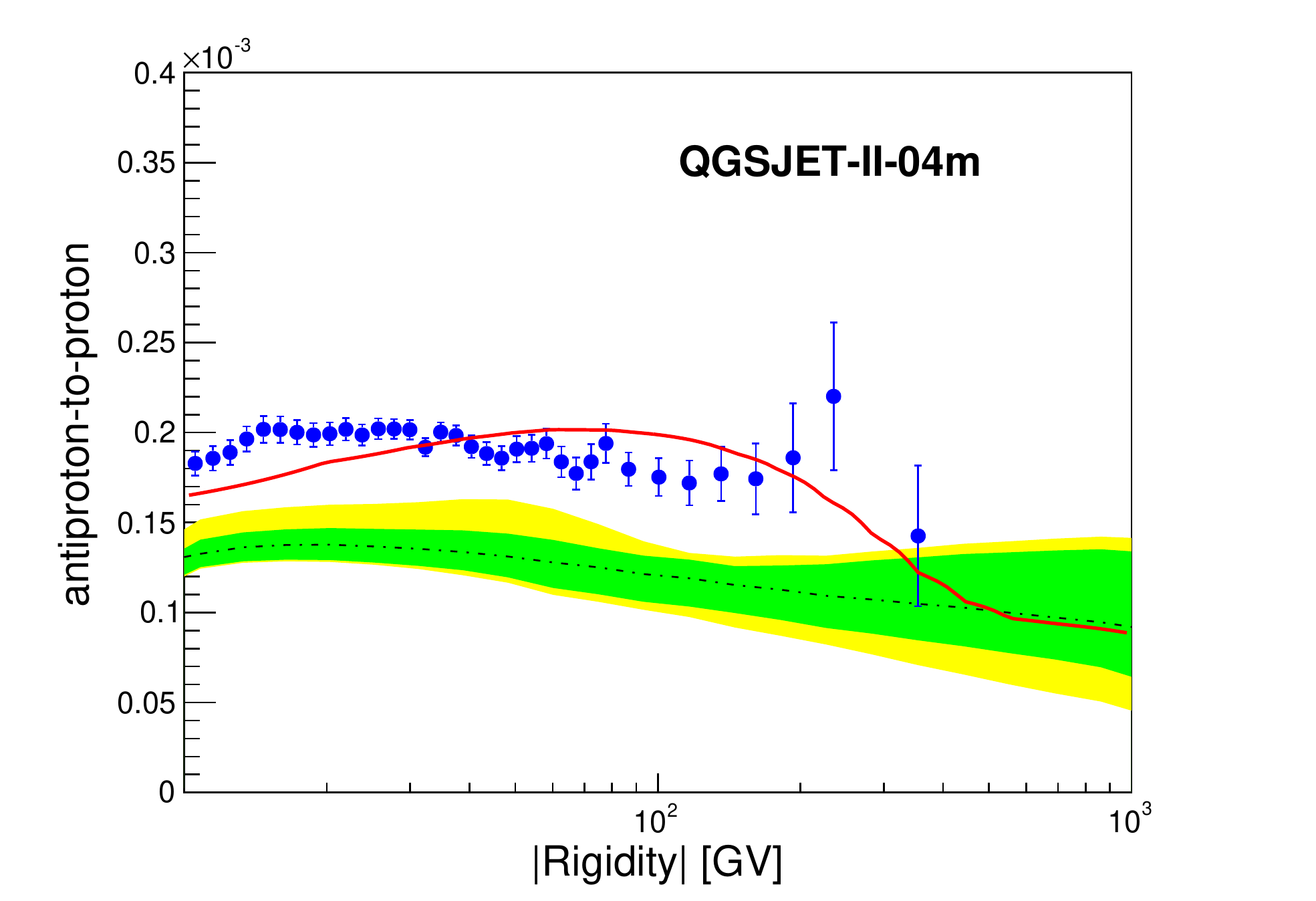}
\includegraphics[width=0.46\textwidth]{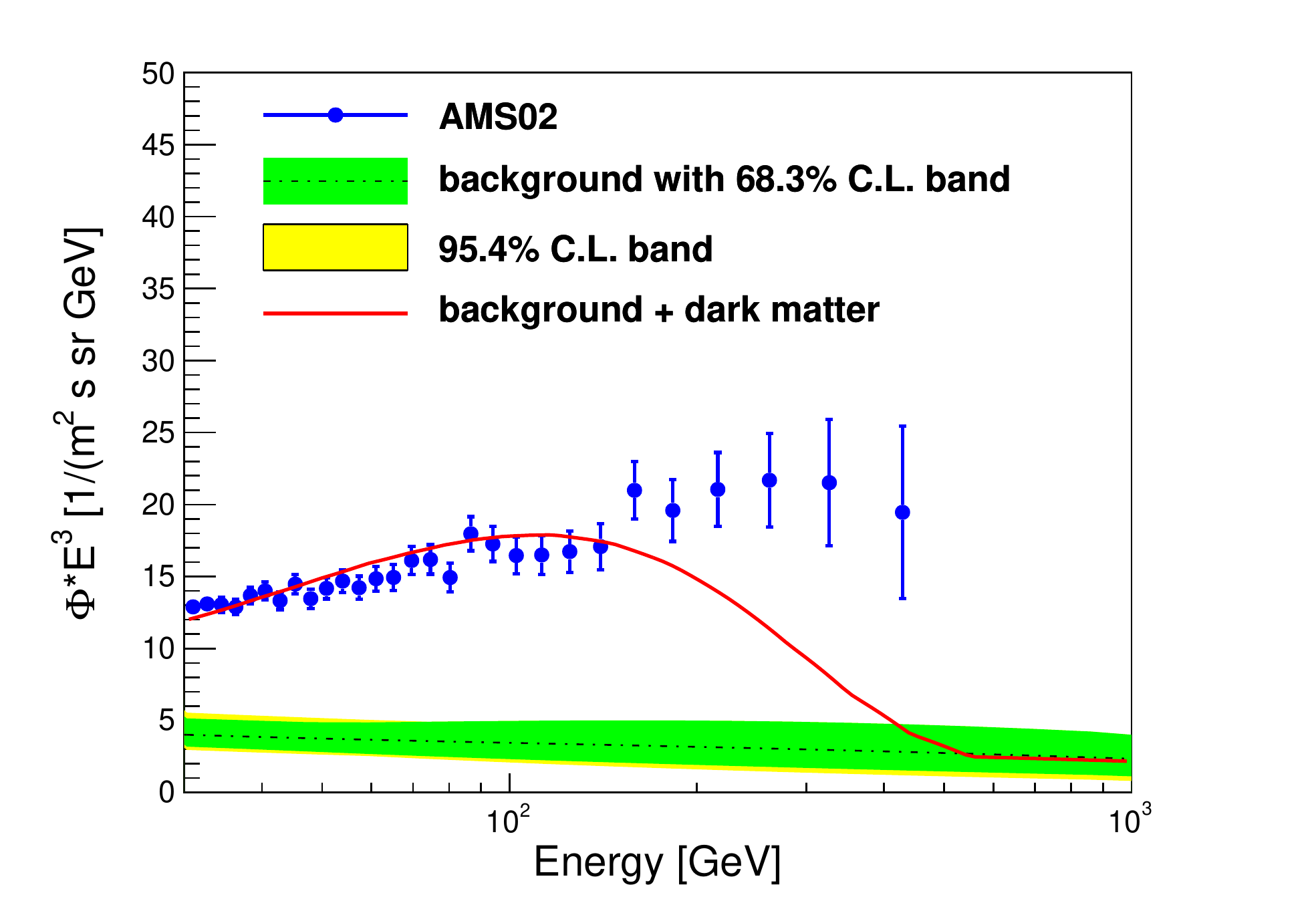}
\caption{
Combined fit of \pbarp{} (TOP) and \eplus{} (BOTTOM) for the 2-channel scenario with QGSJET-II-04m: $\chi\chi\rightarrow\q\qbar$ and $\chi\chi\rightarrow VV \rightarrow2\muplus\muminus$.
}
\label{Fig::fit_2channel_qgsjet}
\end{figure}
%%%%%%%%%%%%%%%%%%%%%%%%%%%%%%%%%%%%%%%%%%%%%%%%%%%%%%%
Now we come to the possibility that dark matter annihilates into two channels.
In Sec.~\ref{Sec::one-channel}, it is shown that more \eplus{} in the annihilation will
improve the fit.
Setting one of the 6 channels in Sec.~\ref{Sec::one-channel} as the first channel, we have studied
``4-body'' lepton channels as the second channels, which are pure lepton
channels and do not produce any \pbar{}. In this kind of scenarios, we will have more \eplus{}
while keeping almost the same amount of \pbar{}.

Seven scenarios with a best fit p-value greater than $10^{-7}$ for EPOS LHC
as the antiproton production model are listed in Table.\,\ref{Tab:results_eposlhc}.
The number of degree of freedom is 208.
\textbf{$CH_i$} stands for the $i$th channel. $BR_i$ is short for the branching ratio of
the $i$th channel.
Compared with the one channel scenarios, these two channel scenarios can improve the quality of the fit
a lot.
In Table.\,\ref{Tab:results_eposlhc}, it is shown that
$\chi\chi\rightarrow VV \rightarrow2\tauplus\tauminus$ is the dominating channel in three scenarios with
the largest p-values.
For QGSJET-II-04m, no scenario gives a p-value greater than $10^{-7}$.
The best scenario gives a $\chi^{2}/n.d.f = 349.11/208$ and a p-value = 3.1$\times 10^{-9}$ with the parameters:
$m_{\chi} = 545$GeV,
\textbf{$\langle\sigma v\rangle\times BR_1$ = $3.45\times10^{-26}cm^3/s$} for $\chi\chi\rightarrow \q\qbar$ and
\textbf{$\langle\sigma v\rangle\times BR_2$ = $74.47\times10^{-26}cm^3/s$} for $\chi\chi\rightarrow VV \rightarrow 2\muplus\muminus$.

We draw the \pbarp{} and \eplus{} plots of dark matter annihilation into \W\Wbar{} and
VV$\rightarrow$2\tauplus\tauminus{} channel as the ``best'' fit example for EPOS LHC in Fig.~\ref{Fig::fit_2channel}.
This scenario shows the mass of dark matter is $1689\pm23$ GeV and its \textbf{$\langle\sigma v\rangle$} = $600.02\pm21.32$
$\times10^{-26}cm^3/s$ with a branching ratio of $0.890\pm0.017 \%$ for $\chi\chi\rightarrow\W\Wbar{}$
and that of $99.11\pm 0.03\%$ for $\chi\chi\rightarrow VV\rightarrow2\tauplus\tauminus{}$.
In Fig.~\ref{Fig::fit_2channel_qgsjet}, it is shown the \pbarp{} and \eplus{} plots of dark matter annihilation into \q\qbar{} and
VV$\rightarrow$2\muplus\muminus{} channel for QGSJET-II-04m.
This fit gives a low p-value, which is 3.1$\times 10^{-9}$. We obtain that the mass of dark matter is $545\pm20$ GeV and its \textbf{$\langle\sigma v\rangle$} = $77.92\pm4.08$
$\times10^{-26}cm^3/s$ with a branching ratio of $4.42\pm0.17\%$ for $\chi\chi\rightarrow\q\qbar{}$
and that of $95.57\pm0.16\%$ for $\chi\chi\rightarrow VV\rightarrow2\muplus\muminus{}$.

These two plots show that the two antiproton production models do not give consistent results for all the scenarios.
As is discussed in Sec.~\ref{Sec::Astro-physical_background}, the difference of
anti-neutron production in these two MC generators is the source of the systematics of the antiproton astro-physical background.
Some recent works parameterized the antiproton production cross sections with the latest ground experimental data
\cite{Kappl:2014hha,Winkler:2017xor,DiMauro:2014iia}, which is also a good way to obtain this cross section.
For antineutron production, however, they assumed an energy independent scale factor $\kappa\equiv$antineutron/antiproton to be
a constant according to isospin symmetry, based on a preliminary experimental result published in a
conference proceeding \cite{Fischer:2003xh}. One should notice that this energy independent assumption of $\kappa$ is not
precise enough to describe antineutron production.
When the antiproton energy is close to the production threshold, $\kappa$ should be maximum in any model.
$\kappa$ goes down when the antiproton energy moves away from the threshold \cite{Feng:2016loc}.
An energy dependent $\kappa$, however, changes the shape of the \pbar{} flux.
More cross section measurement data from accelerators will help to reduce this kind of systematic uncertainties.

As is shown in the top plots of Fig.~\ref{Fig::fit_2channel} and Fig.~\ref{Fig::fit_2channel_qgsjet},
both two MC generators predict a \pbarp{} astrophysical background going down with energy above 60 GeV while AMS-02 data
is flat.
The dark matter signal makes the \pbarp{} spectrum harder and closer to observed data. 
One should notice that these model predictions are based on the spatial-dependent propagation model. 
The standard GALPROP model shows its antiproton astro-physical background calculated with QGSJET-II-04m is compatible with AMS-02 data \cite{Boschini:2017fxq} at high rigidity.
On the other hand, \eplus{} flux measured by AMS-02 is significantly higher than
astrophysical background. 
If the extra source produces the same among the \eplus{} and \eminus{}, one should expect an excess in \eminus{} spectrum and \eplus{} + \eminus{} spectrum. Compared to the astrophysical background, the excess in those spectra \cite{Boschini:2018zdv} is not as significant as that in pure \eplus{} spectrum. 
The dark matter profile can produce a ``cut-off'' like spectrum as is measured by AMS-02.
The ``best'' fit results can match measurement up to a few hundred GeV.

%%%%%%%%%%%%%%%%%%%%%%%%%$$$$$$$$$$$$$$
\section{Conclusions\label{Sec::conclusion}}
%%%%%%%%%%%%%%%%%%%%%%%%%%%%%$$$$$$$$$$
An increase in the accuracy of the CR antiparticle spectra measurements is driving us closer to
the answer of dark matter. Together with CR $\gamma$-ray \cite{FermiLAT:2012ApJ,Ackermann:2015zua,Abdallah:2016jja}
and neutrino \cite{Aartsen:2016pfc} spectra, \pbar{} and \eplus{} spectra
help us study astro-physical properties of the potential dark matter with $m_\chi\sim 10^0-10^5$ GeV.

We summarize everything here.
We present our study on dark matter search from CR \pbar{} and \eplus{} data above 30 GeV.
For the first time, we simultaneously interpretate \pbar{} and \eplus{} spectra in the framework of AMS-02 with dark matter scenarios.
We find that $\chi\chi\rightarrow\tauplus\tauminus{}$ channel with 100\% branching ratio is the best
one channel scenario to reproduce CR \pbarp{} and \eplus{} flux measurement, with $m_\chi$ = $783\pm56$ GeV and
\textbf{$\langle\sigma v\rangle$}=$261.20\pm23.93\times 10^{-26}cm^3/s$, in the case of EPOS LHC.
This scenario is not yet rejected by $\gamma$-ray observation under p-wave cross section assumption.
For the antiproton background using the same MC generator, we also propose a two-channel scenario:
$m_\chi$ = $1689\pm23$ GeV and \textbf{$\langle\sigma v\rangle$} = $600.02\pm21.32$
$\times10^{-26}cm^3/s$.
The dominating channel is $\chi\chi\rightarrow VV\rightarrow2\tauplus\tauminus{}$ with a branching ratio of $99.11\pm0.03\%$,
while the second channel is $\chi\chi\rightarrow\W\Wbar{}$ with a branching ratio of $0.890\pm0.017\%$.
In the case of QGSJET-II-04m, no scenario gives a good fit.
These scenarios predict \pbarp{} spectra harder than those in the background only scenario. They also predict a \eplus{}
spectrum with a cut off between 100 and 2000 GeV, which is also observed by AMS-02, even though the shape does not completely match data. 
Since the direct observation of dark matter annihilation has not yet been reported, 
our results (\ie{} masses, cross sections and channels) can provide useful information for the collider experiments (ATLAS and CMS) to search for weakly interacting massive particles (WIMPs) beyond the standard model.

Comparing to the pulsar scenarios \cite{Feng:2015uta}, we find that the $\chi^2/n.d.f.$s in dark matter fits are higher. This is due to the fact that pulsar
models usually have more degrees of freedom. For example, the injection spectral index of a pulsar is a free parameter, which will adjust the pulsar
profile to match \eplus{} data. On contrast, the spectral index of \eplus{} flux produced by dark matter is fixed by theoretical models.
It is necessary to have some models to constrain the spectral index of a pulsar or to link it with the corresponding $\gamma$-ray spectrum.
Moreover, if one tries to perform a combined fit on \pbarp{} and \eplus{} spectra with quark channel dark matter
and pulsar model, he has 2+3=5 free parameters and will obtain a good fit. Here we have only 3 free parameters for two channel dark matter scenario.
A recent study \cite{DiMauro:2015jxa}  reported  the interpretation of \AMS{} lepton paper with dark matter and pulsar scenarios. 
The solution purposed by the authors of Ref. \cite{DiMauro:2015jxa} contains 3 free parameters for supernova remnants, 2 for pulsars and 2 for dark matter. A fairly good result for a dark matter annihilating in the $\mu^{+}\mu^{-}$ channel was obtained in Ref. \cite{DiMauro:2015jxa}, where its cross section is relatively small and close to the thermal value. Their methods can also be adopted to obtain upper limits for the dark matter scenarios. 
We also investigate the impact of the antiproton background caused by cross sections. Two of the most advanced MC generators,
EPOS LHC and QGSJET-II-04m, do not give consistent results. This disagreement reflects the lack of knowledge of anti-neutron production,
which could be supplemented with new data of future underground experiments.

Recent time dependent \eplus/\eminus{} measurements by PAMELA \cite{Adriani:2016uhu}
confirmed charge-sign-dependent solar modulation models \cite{Maccione:2012cu, Kappl:2015hxv}.
The convectional force field approximation \cite{Gleeson:1968zza} is not precise enough
for us. 
Recent developments of solar modulation model \cite{Bobik:2011ig, Bobik:2015, Boschini:2017gic} 
considered more realistic physical processes. 
This model, namely HelMod, has successfully reproduced proton spectra during solar cycle 23 and 24. 
Another interesting study discovered that solar modulation parameters are related to the number of solar sunspots and the tilt angle of the heliospheric current sheet 8.1 months in advance \cite{Tomassetti:2017gkx}.
AMS-02 will publish its much more precise time-dependent \eplus{}, \eminus{}, \pbar{} and $p$
fluxes in the near future, which will allow 
us to further test HelMod and to reconstruct the fluxes
out of the heliosphere. 
%%%%%%%%%%%%%%%%%%%%%%%%%%%%%%%%%%%%%%%%%%%%
\section*{Acknowledgments}
We thank Qiang Yuan and Zhao-Huan Yu for helpful discussions.
This work is
supported by the National Natural Science Foundation of China (NSFC) under Grant Nos.
11375277, 11410301005, 11647606, and 11005163, the Fundamental Research Funds for the Central
Universities, the Natural Science Foundation of Guangdong Province under Grant No.
2016A030313313, and the Sun Yat-Sen University Science Foundation.
%%%%%%%%%%%%%%%%%%%%%%%%%%%%%%%%%%%%%%%%%%%%%%%%%%%%%%%
\bibliography{appodm}
%%%%%%%%%%%%%%%%%%%%%%%%%%%%%%%%%%%%%%%%%%%%%%%%%%%%%%%
\end{document}